# Integrating formal methods into traditional practices for software development: an overview

Carlos Alberto Fernández-y-Fernández
*Instituto de Computación*
*Universidad Tecnológica de la Mixteca*
*Huajuapan de León, Oax. México*
*caff@mixteco.utm.mx*

*Abstract*—**This paper shows an overview of a research project for integrating formal methods in popular practices for software development in México. The article shows only the main results from the survey about methods and practices and an overview of the initial proposal of practices applying lightweight formal methods to requirements specification and software modelling.**

*Keywords— requirements specification; software modelling; lightweight formal methods.*

## I. INTRODUCTION

Software Engineering has been supporting the use of software development methods such as RUP [1], XP [2], UCD [3] and many others; additionally another set of methods with emphasis in software process management such as SCRUM [4], [5] and TSP [6], [7], among others. In addition we have formal methods, which are a set of mathematical and logic techniques to specify, design, implement and verify information systems [8]. Formal methods are not widely used in development for traditional information systems; in Mexico is even more difficult to find documented success stories [9].

## II. DESCRIPTION OF THE PROJECT

The main aim of the project was to propose the integration of formal methods with popular software modelling and specification practices used in business organisations in order to increase the reliability of the software. Our list of particular goals is presented as follows:

- to identify popular practices used in software development by business organizations,

- to analyse formal methods with possibilities of integration with traditional practices,
- to adapt and design a proposal for the application of formal methods together with a selected set of traditional practices, and
- to evaluate the proposal with software development teams.

Even with the advance in software processes and visual modelling, failures in software development still persist in recent years [10]. One important cause for mistakes in software development is related with a poor specification from the requirements description to the modelling of particular details of the software. It is our opinion that these problems could be reduced if more software specifications and models take advantage of what formal methods offers. However, developers are reticent to use them, therefore we need to present a simpler way to increase their use without adding a whole additional layer of activities to the development cycle.

This research began doing a bibliography review of the techniques and practices of software development to find the more accepted at the moment, making emphasis in the practices that generate a specification or model artifact. In addition, we did a survey among software development companies in Mexico that helped us to select the most accepted practices. Later we reviewed some formal languages, where we put particular attention to the difficulty, being this a factor that may inhibit their acceptance. Then we related the selected practices with the formal methods, trying to propose a set of modifications which integrates and adapts the practices to take into consideration formal aspects. The proposal will be tested with software development teams. 1 shows the methodology mentioned above.

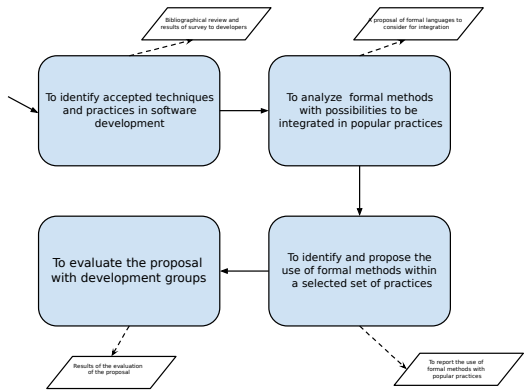

Fig. 1. planned activities

## III. Results on the survey of methods and practices for software developing in México

We applied a survey trying to identify, among other things, the main methods and practices used by software development companies in México. The survey was applied online, which was broadcast to multiple groups, including even with the support of a specialised magazine connecting academia and businesses in software development [11]. The survey was divided into the following areas:

- general questions about methods and practices,
- questions about tools and notations
- question about formal methods and web development

Among the main answers that we wanted to know were precisely about the more popular methods and practices used in Mexico, therefore they are the only data we are presenting here. The whole results of the survey can be consulted in [11].

One of the first things that we wanted to know is if developers in México are using software development methods and in which percentage. 2 shows that only 27% said that they apply some method, while 39% apply a development method but only partially.

In 3 can be seen that, according to our survey, the most popular development method used in Mexican companies is Scrum (22.5%), followed by Extreme Programming, Iterative and incremental approach, and waterfall with almost 12%; the Unified Process and a prototype based approach with almost 9%.

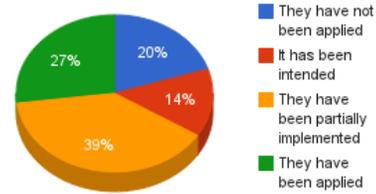

Fig. 2. Do you use any particular software development method?

When trying to identify practices, we could not find conclusive results. We elaborated this particular query as an open question with the idea to offer no influence about any particular practice and being able to receive different practices to the ones widely discussed by us in the academy. The result was, as we mentioned, inconclusive: some of the answers included the mention of a whole method or improvement process such as SCRUM and CMMI, another answers mentioned phases or general activities (e.g., requirements analysis, software design) instead of particular practices. From this we could infer that there is still a poor use or confusion about practices. It is important to mention that the definition for both concepts, method and practice, were included with the survey to reduce the possibility of mistake by the developers. In any case, some of the practices mentioned that called our attention were: pair programming, unit testing, prototyping, and visual modelling. Finally, we think it is important to mention that, even when no particular practice was identified in this area, requirements analysis and specification was the most mentioned by developers. A complete report of the survey can be seen in [11].

## IV. Overview of the proposal: using formal methods with traditional practices

From the survey data but also by analysing the possibilities of particular formal methods we are proposing initially eight practices, four for requirements specification and four for software modelling. We are proposing the use of three lightweight formal methods, attempting to reduce the adoption cost for developers. The practices were divided whether they can or cannot be verified, but considering that the practical output could be already useful even if no yet formally verified. The proposed practices for requirements specifications are as follows:

1. Activity diagram for use case
2. Verification of activity diagrams with task algebra
3. Use case specification with Alloy
4. Verification of use cases with Alloy

And, the proposed practices for software modelling:

1. Modelling class diagrams with Alloy
2. Verifying class diagrams with Alloy
3. Modelling class diagrams with OCL
4. Verifying class diagrams with OCL

Both set of practices were presented using the template proposed in Kuali-Beh by Oktaba [12] and can be consulted in [13].

## V. CONCLUSIONS AND FUTURE WORK

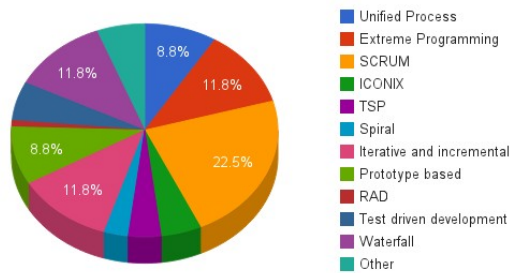

**If using software development methods, indicate which ones**

Fig. 3. If using software development methods, indicate which ones

Based in our survey, there is almost no evidence of use of formal methods in México. We proposed two sets of practices to include lightweight formal methods in requirements specification and software modelling in order to incorporate them easily to the software development cycle.

In the coming months we are expecting to validate our proposal, refining our practices according to the results. Additionally, in the future we would like to include more practices and promote them with software developers.

### ACKNOWLEDGEMENT

This work has been funded by the Universidad Tecnológica de la Mixteca and LANIA. In addition, we would like to acknowledge the support of SG Magazine for spreading our survey.